\documentclass[10pt]{iopart}
\usepackage{iopams} 

\expandafter\let\csname equation*\endcsname\relax
\expandafter\let\csname endequation*\endcsname\relax
\usepackage{amsmath}

\usepackage{graphicx}
\usepackage{bm}
\usepackage{etoolbox}
\usepackage{physics}
\usepackage{stackengine}
\usepackage{mathtools}
\usepackage{lipsum}
\usepackage{breqn}
\usepackage{hyperref}
\usepackage{nicefrac}
\usepackage{dblfloatfix}
\usepackage{afterpage}
\usepackage{amsmath,amsfonts,amssymb}

\usepackage{siunitx}

\usepackage{cuted}

\usepackage{cite}

\def\XXint#1#2#3{{\setbox0=\hbox{$#1{#2#3}{\int}$ }
\vcenter{\hbox{$#2#3$ }}\kern-.6\wd0}}

\newcommand{\Alfven}{Alfv\'{e}n }

\hypersetup{
    colorlinks=true,
    linkcolor=blue,     
    urlcolor=cyan,
    citecolor=blue,
}

\AtBeginEnvironment{align}{\setcounter{subeqn}{0}}
\newcounter{subeqn} \renewcommand{\thesubeqn}{\theequation\alph{subeqn}}%
\newcommand{\subeqn}{%
  \refstepcounter{subeqn}
  \tag{\thesubeqn}
}

\begin{document}

\title[]{Long range frequency chirping of \Alfven eigenmodes}

\author{H. Hezaveh$^1$, Z.S. Qu$^1$, B.N. Breizman$^2$ and M.J. Hole$^{1,3}$}

\address{$^1$ Mathematical Sciences Institute, The Australian National University, Canberra ACT 2601, Australia}
\address{$^2$ Institute for Fusion Studies, The University of Texas at Austin, Austin, TX 78712, USA}
\address{$^3$ Australian Nuclear Science and Technology Organisation, Locked Bag 2001, Kirrawee DC, NSW, 2232, Australia}
\ead{hooman.hezaveh@anu.edu.au}
\date{today}

\begin{abstract}
A theoretical framework has been developed for an NBI scenario to model the hard nonlinear evolution of Global \Alfven Eigenmodes (GAEs) where the adiabatic motion of phase-space sturctures (holes and clumps), associated with the frequency chirping, occurs in generalized phase-space of slowing down energetic particles. The radial profile of the GAE is expanded using finite elements which allows update of the mode structure as the mode frequency chirps. Constants of motion are introduced to track the dynamics of energetic particles during frequency chirping by implementing proper Action-Angle variables and canonical transformations which reduce the dynamics essentially to 1D. Consequently, we specify whether the particles are drifting inward/outward as the frequency deviates from the initial MHD eigenfrequency. Using the principle of least action, we have derived the nonlinear equation describing the evolution of the radial profile by varying the total Lagrangian of the system with respect to the weights of finite elements. For the choice of parameters in this work, it is shown that the peak of the radial profile is shifted and also broadens due to frequency chirping. The time rate of frequency change is also calculated using the energy balance and we show that the adiabatic condition remains valid once it is satisfied. This model clearly illustrates the theoretical treatment to study the long range adiabatic frequency sweeping events observed for \Alfven gap modes in real experiments.        

\end{abstract}

%
%
%
%
\ioptwocol

\section{Introduction}
\label{sec:intro}

\Alfven waves can be unstable as a result of their interaction with energetic particles (EPs) which satisfy the resonance condition during the slowing down process \cite{Chen1994}. In magnetic fusion devices e.g. tokamaks, \Alfven eigenmodes (AEs) \cite{Cheng1985,Heidbrink}, located outside the shear \Alfven continuum, are subject to weak continuum damping and therefore can be destabilized by supra-thermal particles and fusion products. These modes are potentially dangerous for particle transport. The feedback between unstable waves and enhanced particle diffusion would degrade EP confinement (see the review article \cite{Gorelenkovreview} and the references therein and also \cite{Boris2011,Iterphysicsexperts,Darrow}), influence the fuel burnup \cite{Duong,Sigmar,Anderson,Yamagiwa} and subject the material of the containment vessel to increased erosion. On the other hand, destabilization of \Alfven waves may have some beneficial effects e.g. diagnostic purposes of the plasma core \cite{Fasoli2002,Matthew2013}, achieving higher confinement regime due to redistribution of injected ions in DIII-D \cite{Wong2005} and energy channeling of fusion born alpha particles \cite{Herrmann1997}. Therefore, an ability to model and control these kinetically driven instabilities is crucial to the design and operation of a fusion power plant.

The wave-particle interaction, which is essentially one dimensional, may result in frequency sweeping behaviors \cite{HeidbrinkDIIID,Gorelenkov,Wong,sdpinches,Fredrickson2006a,Shinohara}. Refs. \cite{BB,Berk1999}, which successfully explain the chirping observed in experiments \cite{Fasoli1998, Heeter2000}, describe the possible formation of phase--space structures, namely holes and clumps, whose motions are associated with frequency sweeping events. In this model, the radial structure of the MHD mode is fixed, a logical assumption as long as the frequency remains close to the initial eigenfrequency. Subsequently, a nonperturbative model \cite{Boris2010} was presented to investigate the long range sweeping events \cite{Gryaznevich,Maslovsky,Fredrickson2006b} in the hard nonlinear regime where the structure of the mode is considerably affected by the nonthermal fast particles population. Inclusion of collision operators into this model was accomplished in Refs. \cite{Nyqvist2012,Nyqvist2013}. More recently, a 1D theoretical framework was developed in Ref. \cite{Hezaveh2017}, which investigates the impact of different EPs orbit topologies (magnetically trapped/passing) on long range frequency chirping of BGK modes. It should be noted that these models consider the adiabatic evolution of phase--space structures and therefore the Vlasov equation can be bounce averaged to find the perturbed phase--space density of resonant particles. In a very recent work \cite{Wang2018}, a new kinetic code, CHIRP, has been used to study the nonlinear behavior of an energetic particle mode (EPM) which is established and evolved inside the shear \Alfven continuum.

In this paper, we develop a model to describe the hard nonlinear evolution of a Global \Alfven eigenmode (GAE), which is destabilized outside the \Alfven continuum, using a Lagrangian formalism. In a tokamak, the poloidal and toroidal profiles of the mode represent periodic behavior. It was shown in \cite{Boris2010} and \cite{Hezaveh2017} that when the frequency deviates from the initial eigenfrequency significantly, the mode preserves its periodic behavior but does not remain sinusoidal. However, as the first step, we focus on the evolution of the radial structure of the eigenfunction, which does not have a periodic behavior, throughout this paper. Hence, the eigenfunction is presented by a single poloidal $(m)$ and toroidal $(n)$ mode number and the nonlinear contribution of EPs current updates the radial profile of the mode. We retain toroidal effects on EP dynamics in a high aspect ratio tokamak limit: these determine the hard nonlinear evolution of an energetic particle driven mode. The initial eigenfrequency lies just below the shear \Alfven continuum and we study the dynamics associated with a downward branch of frequency chirping. There is no continuum crossing in this case, which is a requirement for the model to remain valid as the frequency chirps. We consider the total Lagrangian of the system and use finite element method to expand the radial structure of the eigenmode. Varying the total Lagrangian with respect to finite element weights gives nonlinear equations describing evolution of the mode radial profile, which is analyzed by invoking the adiabatic condition.  

Section \ref{sec:generalmodel} describes the general picture of the problem in tokamaks. In section \ref{sec:model}, the equation of the mode driven by EPs is presented. The linear growth rate is calculated by finding an explicit expression for perturbed EPs phase--space density. Afterwards, we introduce an adiabatic Hamiltonian describing the dynamics of EPs during frequency chirping, which together with bounce averaging the Vlasov equation, allows us to solve the nonlinear equation for the evolving radial profile of the mode and the rate of frequency chirping. Section \ref{sec:numerical} describes a numerical procedure implemented to solve the nonlinear equations. Section \ref{sec:results} presents the results for a specific polodial and toroidal mode number. This includes equilibrium profiles, dynamics of EPs during frequency chirping, the evolution of the radial profile, the rate of frequency chirping and validation of the adiabatic condition. Section \ref{sec:conclusion} is a summary.

\section{General formalism for nonlinear GAEs}
\label{sec:generalmodel}
We consider a saturated MHD eigenmode with an already established structure in the case of a near-threshold instability. In the presence of weak damping, the coherent group of EPs locked in the mode results in signals with adiabatic frequency chirping in the EPs phase-space associated with the slow evolution of the saturated structure. These signals represent the nonlinear BGK modes with a chirping frequency. In tokamak geometry the general form of a nonlinear chirping mode whose radial profile is evolving slowly/adiabatically can be presented by
\begin{equation}
\Phi \left (\bm{r};t_s;t_f\right )=\sum_{h} \phi_h \left (\bm{r};t_s\right )\e^{-ih\alpha\left (t_f\right ) } + c.c
\label{eq:generalmode}
\end{equation}
with
\begin{eqnarray}
\phi_h=\sum_{m}\phi_{m;n;h}\left (  r; t_s \right )& \e^{ih \left ( m\theta + n\varphi \right )} \nonumber \\
&=\sum_{m,l}\lambda\left (t_s \right ) Y_l \left ( r\right ) \e^{ih \left ( m\theta + n\varphi \right )},
\label{eq:gnrlstr}
\end{eqnarray}
where $m$ and $n$ are the poloidal and toroidal mode numbers, respectively, $ Y_{l} \left (r \right ) $ are base functions and the corresponding weights $\lambda_{l}$ used to describe the finite-element expansion of the radial mode structure, $t_f$ represents fast time scale on the order of the inverse eigenfrequency, $t_s$ represents the slow time scale on which the BGK mode evolves i.e. much longer than the bounce period of the particles trapped in the mode and $\alpha\left (t_f\right )=\int_0^{t_f}\omega\left ( t^{\prime} \right ) dt^{\prime}$. For simplicity, the indices are dropped from $t$ in the following.    

The total Lagrangian describing the system reads
\begin{equation}
L=L_{\text{wave}} + \sum_{\text{fast particles}} L_{\text{particles}} + \sum_{\text{fast particles}} L_{\text{int}},
\label{eq:totallagrangian}
\end{equation}
where $L_{\text{wave}}$ is the MHD wave Lagrangian, $L_{\text{particles}}$ is the EP Lagrangian describing the equilibrium motion and the interaction Lagrangian is denoted by $L_{\text{int}}$. The total Lagrangian contains all the dynamical variables of the system i.e. particle variables and field variables. In principle, one should vary the Lagrangian with respect to each dynamical variable and follow each variable individually. But instead, we use a kinetic description. Nevertheless, each particle trajectory is still a characteristic of the kinetic equation which needs to be solved. However, we circumvent this difficulty by the adiabatic chirping approximation, so we do not solve the kinetic equation directly but we prescribe an approximated shape. Hence, we do not vary the total Lagrangian with respect to fast particle variables but only with respect to the field variables. Starting from the Littlejohns Lagrangian \cite{littlejohn} for the guiding center motion of the fast particles, we have
\begin{equation}
L_{\text{Littlejohn}} = e \left ( \bm{A} + \rho_{\parallel} \bm{B} \right ) \cdot \dot{\bm{X}} + \frac{m_i}{e} \mu \dot{\Omega} - H,
\label{eq:littlejohnlagrangian}
\end{equation}
where $\bm{B}=\nabla \times \bm{A}$ is the total magnetic field with $A$ the vector potential, $\rho_{\parallel}$ is the gyroradius, $m_i$ is the ion mass, e is the electron charge, $\mu$ the magnetic moment, $\dot{\Omega}$ is the time rate of change of gyrophase and $ H =\frac{1}{2} m v_{\parallel}^2 + \mu B$ is the particle Hamiltonian with $v_{\parallel}$ being the particle velocity parallel to the equilibrium magnetic field. It should be noted that we have considered a gauge where the perturbed electrostatic potential is zero. Using Eq. \eqref{eq:littlejohnlagrangian}, the particle and interaction Lagrangian can be found as
\begin{equation}
L_{\text{particles}} = P_{\theta} \dot{\theta} + P_{\varphi} \dot{\varphi} + P_{\Omega} \dot{\Omega} - H_{0} \left ( P_{\theta}, P_{\varphi}, P_{\Omega}, \theta \right )
\label{eq:unperturbedparticlelagrangian}
\end{equation}
and
\begin{equation}
L_{\text{int}} = e \tilde{\bm{A}} \cdot \dot{\bm{X}} = - e \left [ \pdv{\Phi}{t} + v_{\parallel} \bm{b} \cdot \nabla \Phi \right ],
\label{eq:intlagrangian1}
\end{equation}
respectively, where $H_0$ is the equilibrium Hamiltonian written in terms of the canonical variables: canonical momenta $P_\theta$, $P_\varphi$ and $P_\Omega$ conjugated to $\theta$, $\varphi$ and $\Omega$, respectively, $\bm{b}$ is the unit vector in the direction of the equilibrium magnetic field $\bm{B}_0$ and $\tilde{\bm{A}}_{\perp} = \tilde{\bm{A}} - \frac{\bm{B}_0}{B_0^2} \left ( \tilde{\bm{A}} \cdot \bm{B}_0 \right ) $ is the perturbed vector potential that can be represented by two independent scalar functions $\Phi$ and $\Psi$ as
\begin{equation}
\tilde{\bm{A}}_{\perp} = \nabla \Phi - \bm{B}_0 \left ( \bm{B}_0 \cdot \nabla \Phi \right ) / \bm{B}_0^2 + \bm{B}_0 \times \nabla \Psi / \bm{B}_0.
\label{eq:A1}
\end{equation}
As mentioned in \cite{Berk1995}, the compressional perturbation $\Psi$ is almost decoupled from the shear \Alfven perturbation $\Phi$. We, therefore only consider the $\Phi$ term for a GAE.

The periodicity of the unperturbed motion allows us to implement AA variables $(\tilde{\theta},\tilde{\varphi},\tilde{\Omega},P_{\tilde{\theta}},P_{\tilde{\varphi}},P_{\tilde{\Omega}})$ for EPs dynamics. The transformation to AA variables is governed by the following type-2 generating function
\begin{equation}
G_2 = \varphi P_{\tilde{\varphi}} + \Omega  P_{\tilde{\Omega}} + \int_{0}^{\theta} P_{\theta}  d\theta .
\label{eq:AAgenerating}
\end{equation}

The interaction Lagrangian can be written in a more explicit form by substituting Eqs. \eqref{eq:generalmode} and \eqref{eq:gnrlstr} in Eq. \eqref{eq:intlagrangian1}. The perturbed/interaction Hamiltonian reads
\begin{equation}
H_{\text{int}}=-L_{\text{int}}.
\label{eq:deltaH}
\end{equation}
By neglecting the toroidal coupling between poloidal components of the mode and expanding in AA variables of the unperturbed motion, we find   
\begin{eqnarray}
H_{\text{int}} = -\sum_{l,h,p} \lambda_{l,h} \left ( t \right )& V_{p,n,l,h} \left ( P_{\tilde{\theta}},P_{\tilde{\varphi}},P_{\tilde{\Omega}} \right ) \nonumber \\
& \times \e^{ip\tilde{\theta} + ih \left [n \tilde{\varphi} -\alpha \left ( t \right ) \right ]} + c.c,
\label{eq:generaldeltaH}
\end{eqnarray}
where $\tilde{\theta}$ and $\tilde{\varphi}$ are poloidal and toroidal angles, respectively, corresponding to action-angle (AA) variables of the unperturbed motion and $V$ is the coeffcient of Fourier expansion in $\tilde{\theta}$. It should be noted that $p$ is the indice of resonances in the linear case $\left (h=1\right )$ whereas in the nonlinear problem $\frac{p}{h}$ lables different resonances. 

The equations governing the evolution of the mode structure and the frequency can be derived by varying the total lagrangian of the system with resect to the two dynamical field variables, namely $\lambda$ and $\alpha$. We write the system of equations for the evolution of the mode structure by varying the total Lagrangian with respect to $\lambda$, however, in the limit of adiabatic frequency chirping and the slow evolution of the radial profile $\dv{\ln{\lambda_{l}}}{t} \ll \dot{\alpha} $, the energy balance principle is used to track the evolution of the frequency $\left ( \dot{\alpha} \right )$ during chirping. The equation corresponding to each harmonic $\left (h \right )$ of the evolving mode structure reads
\begin{eqnarray}
\left ( h^2\dot{\alpha}^2 \mathsf{M}_h - \mathsf{N}_h \right ) \cdot \bm{\lambda}_h =- \frac{1}{4} \int d^3p d^3q  \delta f  \left ( \bm{q},\bm{p}, t \right )   \nonumber \\
\times\sum_p \begin{bmatrix} V_{p,n,l=1,h} \\ \vdots \\  V_{p,n,l=\textbf{s},h} \end{bmatrix} \e^{ip\tilde{\theta} + ih\left [ n \tilde{\varphi} -\alpha \left ( t \right ) \right ] } + c.c,
\label{eq:generalmodestructure}
\end{eqnarray}
where $\mathsf{M}_h$ and $\mathsf{N}_h$ are matrices corresponding to each harmonic (h) and the integration on the RHS is over the phase-space of the EPs.  It is noteworthy that during the adiabatic evolution, the BGK mode acts like a bucket and the trapped EPs inside this mode will be moved slowly in phase-space, associated with adiabatic frequency chirping. Hence, the perturbed phase-space density $\left (\delta f\right )$ is mainly due to the EPs trapped in the mode and can be found by solving the kinetic equation in the adiabatic regime. 

In what follows, the theoretical picture is developed for a single harmonic $\left (h=1 \right )$. In other words, this model focuses on the evolution of the radial component of the mode during adiabatic frequency chirping.  
\section{The model}
\label{sec:model}

We consider a GAE mode in the following calculations and therefore retain only one poloidal harmonic in the linear response of the cold particles. We, however, take into account poloidal variation of the confining magnetic field in our calculations of the EP trajectories and use a high aspect ratio approximation for that.

\subsection{MHD wave Lagrangian}
\label{subsec:MHDwaveLagrangian}
The kinetic ($K$) and potential energy ($W$) of MHD waves are given by
\begin{align}
&K=\frac{1}{2} \int \rho \dot{\bm{\xi}}^2 dV \label{eq:1a} \refstepcounter{equation}\subeqn \\
&W=-\frac{1}{2} \int \bm{\xi} \cdot \bm{F} \left (\bm{\xi} \right ) dV ,\label{eq:1b}   \subeqn 
\end{align}
where $\rho$ is the mass density, $\bm{\xi}$ the displacement vector, $\bm{F}$ is the force operator, $dV$ denotes the differential volume element and the integration is performed over the whole plasma volume. For a low-$\beta$ system and the linearized force operator and in the aforementioned gauge where the perturbed scalar electrostatic potential is zero, the wave Lagrangian reads
\begin{equation}
L_{\text{w}} = \frac{1}{2} \int \rho \left (\frac{\dot{\tilde{A}}_{\perp}}{B_0}\right )^2 - \frac { \abs{\nabla \times \tilde{A}_{\perp}} ^2}{\mu_{0}} + \frac{J_{\parallel}}{B_0} \tilde{A}_{\perp} \cdot \nabla \times \tilde{A}_{\perp} dV  
\label{eq:wlagrangian1}
\end{equation}
where $\tilde{A}$ is the perturbed vector potential, $B_0$ is the equilibrium magnetic field, $\mu_{0}$ is the magnetic permeability, $J_{\parallel}$ is the unperturbed plasma current parallel to the equilibrium field and we have neglected the nonlinear bulk plasma response. Considering only the $\Phi$ term in Eq. \eqref{eq:A1} for a GAE, Eq. \eqref{eq:wlagrangian1} reduces to
\begin{eqnarray}
L_{\text{w}} = \frac{1}{2 \mu_0} \int  \frac{\mu_0 \rho}{B_0^2} \left ( \nabla_{\perp} \dot{\Phi} \right ) ^ 2 - \left ( B_0 \nabla_{\perp} \frac{ \bm{B}_0 \cdot \nabla \Phi }{B_0^2}    \right )^2 - \nonumber \\  
\left [ \frac{\left (  \nabla \times \bm{B}_0 \right ) \left ( \bm{B}_0 \cdot \nabla \Phi \right )}{B_0^2}  \right ] ^2 - \frac{\left ( \bm{B}_0 \cdot \nabla \Phi \right ) \left ( \nabla \Phi \cdot \Delta \bm{B}_0 \right )}{B_0^2} dV,
\label{eq:wlagrangianreduced}
\end{eqnarray}
which can be varied with respect to $\Phi$ to obtain the linear dispersion relation. For a single global \Alfven eigenmode, $\Phi$ can be written as
\begin{eqnarray}
\Phi \left ( \bm{r}; t\right )=\sum_{l=1}^{\textbf{s}} \lambda_{l} \left (t \right ) Y_{l} \left ( r \right ) \e^{im\theta + in\varphi - i \alpha \left ( t\right )} + c.c,
\label{eq:phi1}
\end{eqnarray}
where $\left (r,\theta,\varphi \right )$ are cylinderical coordinates, $\alpha \left (t \right )$ represents rapid oscillations and $\textbf{s}$ is the total number of finite elements. It should be noted that $\lambda_{l}$ and ${\alpha}$ are real quantities and $\lambda_{l}$ is assumed to change slowly compared to the mode frequency, $\dv{\ln{\lambda_{l}}}{t} \ll \dot{\alpha}$. This implies a proper set of the base functions that can represent a smooth radial profile of the global eigenmode. Substituting Eq. \eqref{eq:phi1} into Eq. \eqref{eq:wlagrangianreduced} gives
\begin{equation}
L_{\text{wave}} = 2 \dot{\alpha}^{2} \left [\bm{\lambda}^{\intercal} \cdot \mathsf{M} \cdot \bm{\lambda} \right ] - 2 \left [ \bm{\lambda}^{\intercal} \cdot \mathsf{N} \cdot \bm{\lambda} \right ]    ,
\label{eq:reducelagrangian}
\end{equation}
where the fast time varying part is integrated out and the superscript $\intercal$ denotes the transpose operation, \{$\bm{\lambda}$\}$ \in \mathbb{R}^{\textbf{s} \times 1}$, \{$\mathsf{M},\mathsf{N}$\} $ \in \mathbb{R}^{\textbf{s} \times \textbf{s}}$ whose elements are given by
\begin{subequations}
\begin{eqnarray}
M_{j,k} = \frac{1}{2\mu_0}   \int \frac{r R_0}{V_A^2} \left [ \dv{Y_{j}}{r} \dv{Y_{k}}{r} + \frac{m^2}{r^2} Y_{j} Y_{k} \right ] dr d \theta d \varphi  \label{eq:M}   \\ 
N_{j,k} =  \frac{1}{2\mu_0}   \int  \left [ B_0^2 \left ( \dv{}{r}\frac{k_{\parallel} Y_{j}}{B_0}  \right ) \left ( \dv{}{r}  \frac{k_{\parallel} Y_{k}}{B_0}  \right ) + \left ( \frac{m^2 k_{\parallel}^2}{r^2}    \right. \right. \nonumber \\ 
\left. \left. + \frac{\mu_0^2 J_{\parallel}^2 k_{\parallel}^2}{B_0^2} + \mu_0 m k_{\parallel} \dv{}{r} \left (  \frac{J_{\parallel}}{B_0}  \right )  \right ) Y_{j} Y_{k} \right ] r R_0 dr d\theta d\varphi,  \label{eq:N} 
\end{eqnarray}
\end{subequations}
where $R_0$ is the major radius, $V_A$ is the \Alfven velocity and $k_{\parallel}$ is the wavenumber parallel to the equilibrium magnetic field.

\subsection{Energetic particle and interaction Lagrangian}
\label{subsec:partintlag}

We write the unperturbed particle Lagrangian part of Eq. \eqref{eq:littlejohnlagrangian} using the high aspect ratio tokamak limit where the flux surfaces are approximated by the contours of constant $r$ \cite{Abel2009}. In these coordinates, one can write 
\begin{equation}
 \bm{A}_0 = \psi \nabla \theta - \chi \nabla \varphi + \nabla \eta, \label{eq:AA}
\end{equation}
and
\begin{equation}
 \bm{B}_0 = \nabla \psi \times \nabla \theta - \nabla \chi \times \nabla \varphi 
        = B_{\theta} \nabla \theta + B_{\varphi} \nabla \varphi. \label{eq:BB}
\end{equation}
where $\bm{A}_0$ is the equilibrium part of the vector potential and $\chi$ is the poloidal flux. We have $\nabla \varphi = R^{-1}\bm{e}_{\varphi} $ and according to Amperes law $B_0 \propto R^{-1}$. Hence, $B_{\varphi} \approx B_0R_0$ is a constant with $B_0$ being the equilibrium magnetic field at the center of the plasma. Using Eqs. \eqref{eq:AA} and \eqref{eq:BB}, we have
\begin{equation}
\pdv{\psi}{r} \approx rB_0 \left (1 - \epsilon \cos \theta \right ),
\label{eq:psi}
\end{equation}
where $\epsilon$ is the inverse aspect ratio.

Using a proper gauge to cancel $\nabla \eta$ in $\bm{A}_0$, $L_{\text{particles}}$ can be written in the canonical form given by Eq. \eqref{eq:unperturbedparticlelagrangian} and the canonical variables are given by
\begin{align}
 P_\theta =& e \psi(r, \theta) + m_i v_\parallel b_\theta(r,\theta), \label{eq:p_theta}\\
 P_\varphi =& - e \chi(r) + m_i v_\parallel b_\varphi(r,\theta), \label{eq:p_varphi} \\
 P_\Omega =& \frac{m_i}{e} \mu,
\end{align}
where $b_\theta$ and $b_\varphi$ are covariant components of $\bm{b}$, with $b_\theta \approx r^2/qR$ and $ b_\varphi \approx R$, and $\chi \left (r \right )$ can be found using Eq. \eqref{eq:psi} and the safety factor $q \left (r,\theta \right )=\frac{\bm{B}_0\cdot \nabla \varphi}{\bm{B}_0\cdot \nabla \theta}=\frac{\partial \psi/\partial r}{\partial \chi / \partial r} \approx q\left (r  \right )$. 
The conversion from $r$ and $v_\parallel$ to the canonical variables are now implicitly given by Eq. \eqref{eq:p_theta} and Eq. \eqref{eq:p_varphi}.

The large aspect ratio assumption allows us to drop the $\theta$ dependency in Eq. \eqref{eq:psi}. Taking into account that the toroidal equilibrium field is dominant, $b_\theta$ can also be neglected. Hence, we have
\begin{align}
 P_\theta =&  \frac{1}{2} e X_r^2 B_0, \label{eq:ptheta2} \\
 P_\varphi =& - e \chi \left ( X_r \left ( P_{\theta} \right ) \right ) + m_i v_\parallel R, \label{eq:pphi2}
\end{align}
and
\begin{equation}
 H_0=\frac{1}{2} m_i v_{\parallel}^2 \left ( P_{\theta}, P_{\varphi}, P_{\Omega}, \theta \right ) + \mu B_0 \left [ 1- \frac{X_r \left ( P_{\theta} \right )}{R_0} \cos \theta \right ],
\label{eq:}
\end{equation}
where $X_r$ is the radial position of the EPs. By implementing the canonical equations of motion, we find
\begin{align}
\pdv{H_0}{P_\theta} =&  m_i v_{\parallel} \pdv{H_0}{P_\theta} - \mu B_0 \frac{1}{R_0} \cos \theta \dv{X_r}{P_\theta} = \dot{\theta} , \label{eq:canon1}\\
\pdv{H_0}{P_\varphi} =&  m_i v_{\parallel} \pdv{v_{\parallel}}{P_\varphi} = \dot{\varphi} , \label{eq:canon2}\\
\pdv{H_0}{\theta} =&    m_i v_{\parallel}   \pdv{v_{\parallel}}{\theta} + \mu B_0 \frac{X_r}{R_0} \sin \theta = - \dot{P}_{\theta}. \label{eq:canon3} 
\end{align}
Using Eq. \eqref{eq:pphi2}, we get
\begin{align}
\dot{P}_{\theta} =&  - \left [ \frac{m_i v_{\parallel}^2 }{R} +\frac{\mu B_0}{R_0} \right ] X_r \sin \theta, \label{eq:pthetadot}\\
\dot{\varphi} =&  \frac{v_{\parallel} }{R} , \label{eq:phidot}\\
\dot{\theta} =&   \frac{v_{\parallel}}{R q(X_r)} - \left [ \frac{m_i v_{\parallel}^2 }{R} +\frac{\mu B_0}{R_0} \right ] \cos \theta \frac{1}{e B_0 X_r}. \label{eq:thetadot} 
\end{align}
It should be mentioned that in the high aspect ratio tokamak limit, the safety factor can be considered only as a function of the radius.

The conjugate momenta corresponding to $\varphi$ and $\Omega$ are constants of motion. Therefore, in terms of AA variables introduced by \eqref{eq:AAgenerating}, one can set $P_{\tilde{\varphi}} = P_\varphi$, $ P_{\tilde{\Omega}} = {P}_\Omega $ and considering the definition of the angle variables from $0$ to $2\pi$, we find $\tilde{\varphi}= \varphi + \Delta \varphi$ for motion in the direction of the field line. In this work, we consider the case of a neutral beam injection (NBI) where the majority of the EPs are deeply passing $\left (\mu \approx 0\right )$ inside the equilibrium field. Consequently, $v_{\parallel}$ becomes a constant of motion. We also assume the maximum orbit width $ \left ( \Delta r  \right ) $ to be much smaller than the width of the radial mode structure and let $ r_0 \left ( P_{\theta}, P_{\varphi}, P_{\Omega} \right ) $ be the average position of a drift orbit. These conditions together with the large aspect ratio assumption make $\varphi$ and $\theta$ approximately linear in time. Therefore, we find
\begin{align}
\dot{\tilde{\theta}}= \omega_{\tilde{\theta}} \approx \frac{V_{\parallel}\left ( \tilde{P}_\theta,\tilde{P}_{\varphi},\tilde{P}_{\Omega}  \right ) }{q\left (r_0 \right ) R_0 }	\refstepcounter{equation} \label{eq:za} \subeqn \\
\dot{\tilde{\varphi}}=\omega_{\tilde{\varphi}} \approx \frac{V_{\parallel}\left ( \tilde{P}_\theta,\tilde{P}_{\varphi},\tilde{P}_{\Omega}  \right ) }{R_0}, \label{eq:za} \subeqn
\end{align}
where $\omega_{\tilde{\theta}}$ and $\omega_{\tilde{\varphi}}$ represent the poloidal and toroidal guiding center frequency of deeply passing EPs.

Substituting Eq. \eqref{eq:phi1} into Eq. \eqref{eq:intlagrangian1} results in
\begin{eqnarray}
L_{\text{int}} =& ie \left [ \dot{\alpha} - v_{\parallel} \frac{\left ( \frac{m}{q \left (X_{r} \right )} + n  \right )}{R_0} \right ]  \nonumber \\
&\times \sum_{l} \lambda_{l} \left ( t \right ) Y_{l} \left ( X_{r} \right ) \e^{im\theta +in \varphi -i \alpha \left ( t \right )} + c.c.
\label{eq:intlgr}
\end{eqnarray}
Now, we express the above Lagrangian in terms of the action-angle variables of the unperturbed motion to find
\begin{equation}
L_{\text{int}}= \sum_{l} \sum_{p} \lambda_l \left ( t \right ) V_{p,n,l} \e^{ip\tilde{\theta} + in \tilde{\varphi} -i\alpha \left ( t \right ) } + c.c,
\label{eq:}
\end{equation}
where the coupling strength, $V_{p,n,l}$, is determined by
\begin{eqnarray}
V_{p,n,l} =& \frac{1}{2 \pi} \int ie \left [ \dot{\alpha} \left ( t \right ) - v_{\parallel} \frac{\left ( \frac{m}{q \left (X_{r} \right )} + n  \right )}{R_0} \right ] \nonumber \\ 
& \times Y_{l} \left ( X_{r} \right ) \e^{im \theta + in \varphi} \e^{-ip\tilde{\theta}-in\tilde{\varphi}} d \tilde{\theta},
\label{eq:V_p}
\end{eqnarray}
whose detailed calculation for deeply co-passing orbit types of the EPs is presented in \ref{App1}.

\subsection{Mode equation}
\label{subsec:mode_eq}

The total Lagrangian for one eigenmode can be presented by
\begin{eqnarray}
L &=  2 \dot{\alpha}^{2} \left [ \bm{\lambda}^{\intercal}  \mathsf{M}  \bm{\lambda} \right ] - 2 \left [ \bm{\lambda}^{\intercal} \mathsf{N}  \bm{\lambda} \right ] \nonumber \\
&+ \sum_{\text{fast particles}} P_{\tilde{\theta}} \dot{\tilde{\theta}} + P_{\tilde{\varphi}} \dot{\tilde{\varphi}} + P_{\tilde{\Omega}} \dot{\tilde{\Omega}} - H_{0} \left (P_{\tilde{\theta}}, P_{\tilde{\varphi}}, P_{\tilde{\Omega}} \right ) \nonumber \\
&+ \sum_{\text{fast particles}} \bm{\lambda}^{\intercal} \left (t \right ) \bm{D} +c.c,
\label{eq:89}
\end{eqnarray}
where the elements of \{$\bm{D}$\} $ \in \mathbb{R}^{S \times 1}$ are $D_{l,1}=\sum_{p}  V_{p,n,l} \e^{ip\tilde{\theta} + in \tilde{\varphi} -i\alpha \left ( t \right )  } $. Varying the above Lagrangian with respect to $\bm{\lambda}$ gives the following expression for the nonlinear mode structure 
\begin{eqnarray}
\left ( 4 \dot{\alpha}^2 \mathsf{M} - 4 \mathsf{N} \right )  \bm{\lambda} + \sum_{\text{fast particles}} \bm{D} + c.c =0
\end{eqnarray}
The sum over fast particles can be replaced by integration over initial phase--space. The canonicity of the transformation from the initial phase--space coordinates to the instant coordinates $   \left [  \left (  \bm{q}_0 , \bm{p}_0  \right ) \rightarrow  \left (  \bm{q} , \bm{p} \right )   \right ]  $ allows us to write 
the phase-space integration in terms of the instant coordinates. In addition, as mentioned in section \ref{sec:generalmodel}, we take $h=1$ and therefore we find the nonlinear mode equation in the form given by Eq. \eqref{eq:generalmodestructure} with $h=1$, where $\delta f=\sum_p \delta f_p$ is the perturbed part of the total distribution function $(f=\delta f+F_0 )$ of the EPs with $F_0$ being the equilibrium part. In this model, krook type collisions inside the bulk plasma provide damping mechanism to the wave amplitude at a rate $\gamma_d$, which is implicitly included in Eq. \eqref{eq:generalmodestructure} (see \cite{Berk1995BOT} and \cite{Berk1995}). 

\subsection{Mode evolution}
\label{subsec:}

The total Hamiltonian of the EPs during the hard-nonlinear evolution reads,
\begin{equation}
H=H_{0} \left ( P_{\tilde{\theta}}, P_{\tilde{\varphi}}, P_{\tilde{\Omega}} \right ) + H_{\text{int}}
\label{eq:totalhamiltonian}
\end{equation}
where $H_{\text{int}}=-L_{\text{int}}=\sum_p H_{\text{int},p}$ written in terms of the action-angle variables of the unperturbed motion. In order to simplify the dynamics, we consider the canonical transformation using the type-2 generating function
\begin{equation}
F_2 \left ( \bm{q},\bm{p_{\text{new}}} ,t \right ) = P_{1} \left [ p \tilde{\theta} + n \tilde{\varphi} - \alpha \left ( t \right ) \right ] + P_{2} \tilde{\varphi} + P_{3} \tilde{\Omega},
\label{eq:gf2}
\end{equation}
for the $p$-th resonance. The new variables are defined as follows
\begin{equation}
\begin{split}
P_1 &= \frac{1}{p} P_{\tilde{\theta}}\\ P_2 &= P_{\tilde{\varphi}} - \frac{n}{p} P_{\tilde{\theta}}\\  P_3 &= P_{\tilde{\Omega}}
  \end{split}
\ \ \ \ \ \
\begin{split}
Q_1 &= \zeta = p \tilde{\theta} + n \tilde{\varphi} - \alpha \left ( t \right )\\ Q_2 &= \tilde{\varphi}\\ Q_3 &= \tilde{\Omega} 
 \end{split}
\label{eq:CT1}
\end{equation}
which shows that the wave-particle interaction is effectively one-dimensional in an isolated resonance, i.e. $P_2$ and $P_3$ corresponding to ignorable coordinates, are constants of the motion. It should be emphasized that the above canonical transformation is defined for a specific value of $p$ corresponding to the $p$-th resonance. This can be emphasized by considering a subscript $p$ on the new variables. However, such subscripts are neglected for simplicity. 

In what follows, we first calculate an analytic expression for the perturbed phase--space density of EPs in the linear limit. Afterwards, the dynamics of the resonant particles during the adiabatic chirping of the GAE are identified, followed by the perturbed distribution function during the evolution of holes/clumps.

\subsubsection{Linear regime}
\label{subsubsec:Linear}

The linearized Vlasov equation for the $p$-th resonance
\begin{equation}
\pdv{\delta f_p}{t} + \pdv{\delta f_p}{\zeta} \pdv{ H_{0}}{P_1} =\left. \pdv{F_0}{P_1} \pdv{H_{\text{int},p}}{\zeta} \right |_{P_2, P_3},
\label{eq:Vlasov}
\end{equation}
\sloppy where $\delta f_p=\hat{f}_p\left (P_1 \right ) \exp(i\zeta)+ c.c $ and $H_{\text{int},p}=- \sum_{l} \lambda_l  V_{p,n,l} \exp (i\zeta) + c.c $, can be used to derive an analytic expression for $\delta f_p$. It should be noted that during the linear evolution, we set $\alpha (t)=\omega t$, with $\omega$ being the complex frequency have the real part $\omega_r$ and imaginary part $\gamma_l$. By substituting the relevant expressions in Eq. \eqref{eq:Vlasov}, we find
\begin{align}
\hat{f}_p = \frac{\sum_l \lambda_l V_{p,n,l}  (\pdv{F_0}{P_{\tilde{\varphi}}}n + \pdv{F_0}{P_{\tilde{\theta}}}p )}{\omega -p \omega_{\tilde{\theta}} - n \omega_{\tilde{\varphi}}},
\label{eq:fp}
\end{align}
which gives the resonance condition
\begin{equation}
\omega_{r} = p \omega_{\tilde{\theta}} + n \omega_{\tilde{\varphi}}.
\label{eq:resonance}
\end{equation}

In the limit of deeply passing particles, Eqs. \eqref{eq:generalmodestructure} (with $h=1$) and \eqref{eq:fp} are used to find the linear dispersion relation of the mode given by
\begin{eqnarray}
\left (\omega^2 \mathsf{M} - \mathsf{N}\right)& \bm{\lambda} = \nonumber \\
& 4\pi^3  \sum_p \int dP_1 dP_2   \frac{\left. \pdv{F_0}{P_1}\right |_{P_2, P_3} }{ G \left (P_1 \right ) - \omega }    \bm{T} \bm{\lambda},
\label{eq:linearmode}
\end{eqnarray}
where $G \left (P_1 \right ) = \left. \pdv{H_0}{P_1} \right |_{P_2, P_3} $, \{$\mathsf{T}$\} $ \in \mathbb{C}^{S \times S}$ whose elements are given by $T_{j,k} = V_{p,n,l=j} V_{p,n, l=k}^*$ and $\bm{\lambda}$ represents the initial MHD eigenvector. Neglecting the  infinitesimal contribution from the principal value allows us to set $\omega_r = \omega_{\text{GAE}}$, where $\omega_{\text{GAE}}$ is the initial MHD frequency of the eigenmode. Therefore, the linear growth rate of the mode is found to be
\begin{eqnarray}
\gamma_l = & \left [ \sum_p \int dP_2 \left. \pdv{F_0}{P_1}\right |_{P_2, P_3}  \left (\pdv{G}{P_1} \right )^{-1}  \right. \nonumber \\    
& \times \left.  \bm{T} \left. \bm{\lambda} \right |_{P_1 = P_{1,\text{res}}}  \right ]      \frac{ 2\pi^4 \bm{\lambda}^{\intercal}}{\omega_{\text{GAE}} \bm{\lambda}^{\intercal} \mathsf{M} \bm{\lambda}},
\label{eq:lgrowth}
\end{eqnarray}
where $P_{1,\text{res}}$ is the value of $P_1$ at resonance denoted by $\Pi$ throughout this manuscript and we have assumed $\gamma_l \ll \omega_{\text{GAE}}$. 

\subsubsection{Nonlinear chirping GAE}
\label{subsubsec:Chirping}

The existence of the damping mechanism introduced in subsection \ref{subsec:mode_eq} leads to an unstable plateau in the phase--space density of EPs which supports sideband oscillations that evolve into chirping modes \cite{BB,lilley2015}. For the purpose of investigating the mode during frequency sweeping, we consider a marginal instability case where mode overlap is neglected and we take the limit where phase--space structures (holes and clumps) move adiabatically. Hence, we have
\begin{equation}
\left [\dv{\omega_{b}}{t}, \dv{\dot{\alpha}}{t} \right ]\ll \omega_{b}^{2} \sim \gamma_l^2 \sim \gamma_{d}^2,
\label{eq:adb_condition}
\end{equation}
where $\omega_{b}$ is the bounce frequency of EPs trapped inside the separatrix. Therefore, the finite element amplitudes, $\lambda_l(t),$ evolve on a slow time scale; however, $ \alpha(t)$ includes a fast time scale on the order of $\omega_{\text{GAE}}^{-1}$, which coresponds to the periodic behavior of the field. The canonical transformation presented by Eq. \eqref{eq:CT1} can be implemented to cancel this fast time scale dependency from the Hamiltonian given by Eq. \eqref{eq:totalhamiltonian}. Therefore, for the $p$-th resonance, the total Hamiltonian converts to
\begin{eqnarray}
K= H_0 \left (P_1, P_2, P_3 \right ) - \dot{\alpha} P_1  \\ \nonumber
-\left [ \sum_{p^{\prime}=p,l} \lambda_l V_{p^{\prime}, n, l} \left ( P_1, P_2, P_3  \right ) \e^{i\zeta} + c.c \right ],
\label{eq:newhamiltonian1}
\end{eqnarray}
where highly oscillating terms corresponding to other resonances ($p^{\prime} \neq p$) have been neglected. Assuming the separatrix width to be small compared with the characteristic width of the distribution function, we can Taylor expand the quantities around the middle of the separatrix $\left ( \Pi(t) \right )$, so we have
\begin{eqnarray}
K \approx H_0 \left (\Pi, P_2, P_3 \right ) + \pdv{H_0}{P_1} \left (\Pi, P_2, P_3 \right ) \left [ P_1 - \Pi  \right ]- \dot{\alpha} P_1 \nonumber \\
+ \frac{1}{2}\pdv[2]{H_0}{P_1} \left [ P_1 - \Pi  \right ]^2  - \sum_{l} \lambda_l V_{p, n, l} \left ( \Pi, P_2, P_3  \right ) \e^{i\zeta_p}  + c.c,
\label{eq:newhamiltonian2}
\end{eqnarray}
It should be mentioned that the higher order terms in the expansion of equilibrium Hamiltonian have been neglected due to the smallness of the separatrix width. $\Pi$ satisfies: $\pdv{H_0}{P_1}\left ( \Pi, P_2, P_3 \right ) = \dot{\alpha} \left ( t \right )$, consequently $\Pi=\Pi \left ( P_2, P_3, t \right )$. Therefore, the new Hamiltonian is
\begin{eqnarray}
K \approx & \frac{1}{2}\pdv[2]{H_0}{P_1} \left (\Pi, P_2, P_3 \right ) \left [ P_1 - \Pi  \right ]^2      \nonumber \\
& - \sum_{l} \lambda_l V_{p, n, l} \left ( \Pi, P_2, P_3 \right ) \e^{i\zeta_p}  + c.c .
\label{eq:newhamiltonian3}
\end{eqnarray}
which evolves adiabatically during the frequency sweeping. It is noteworthy to mention that for $\pdv[2]{H_0}{P_1} \left (\Pi, P_2, P_3 \right ) \textgreater 0$, substituting $K$ in Eq. \eqref{eq:newhamiltonian3} with maximum value of $\sum_{l} \lambda_l V_{p, n, l} \left ( \Pi, P_2, P_3 \right ) \e^{i\zeta_p} + c.c$ gives the dynamics on the separatrix. The preserved adiabatic invariant corresponding to the above slowly evolving Hamiltonian is
\begin{equation}
I = \frac{1}{2 \pi} \int P_1 d \zeta
\label{eq:adiabaticinvariant}
\end{equation}
and we denote the corresponding angle by $\eta$. The above equation can be solved for each $P_2$, corresponding to a separatrix, by substituting for $P_1$ and integrating from $0$ to $2\pi$ over the angle variable $\zeta$.

The perturbed distribution of the passing particles traveling around the separatrix remains approximately close to the equilibrium distribution \cite{Boris2010,Nyqvist2012}. Hence, the perturbed density is assumed to be dominantly from the trapped particles inside the separatrix. Considering the small separatrix width assumption mentioned above and followed by bounce-averaging the Vlasov equation (see section 3 in \cite{Nyqvist2012} and appendix B in \cite{Hezaveh2017}), we find
\begin{eqnarray}
\label{eq:perturbedpart}
   \delta f =
    \begin{cases}
      f_0 - F_0(t) = F_0(t=0) - F_0(t) , & $trapped$\\
      0, & $passing$
    \end{cases}
\label{eq:deltaf}
\end{eqnarray}
where $\delta f$ is the perturbed distribution function of the particles inside holes/clumps, $f_0$ is the lowest order term in the expansion of $f$ around the small parameter $\eta=\frac{\tau_b}{\tau_s}$ with $\tau_b$ and $\tau_s$ being the bounce period and the slow time scale of mode evolution, respectively. It should be noted that $t=0$ denotes the initial stage of chirping in this paper. However, for the case of an expanding separatrix and for newly trapped particles during chirping, $t=0$ in the above expression implies the time when EPs are trapped inside the separatrix. 




\subsubsection{Chirping rate}
\label{subsubsec:rate}

According to \cite{Berk1996}, the dissipated power $(Q)$ via weak collisions due to the work of friction force is $2 \gamma_{d}E_{\text{wave}}$, with $E_{\text{wave}}$ being the MHD energy of the mode, which consists the perturbed energy of the cold plasma and the perturbed electromagnetic field. This absorbed power $(Q)$ should be equal to the power $(P)$ released by the phase--space structures energy. Therefore, we have
\begin{equation}
2 \gamma_{d}E_{\text{wave}}= - \sum_{P_2} N_{P_2} \dv{E}{t},
\label{eq:powerbalancecondition1}
\end{equation}
where $N_{P_2}$ is the perturbed number of EPs inside each coherent phase--space structure (hole/clump) in the interval $\Delta P_2$, given by
\begin{equation}
N_{P_2} = \iint \delta f dP_{1} d\zeta \Delta P_2,
\end{equation}
$E$ is the energy of each fast particle inside the hole/clump. This energy consists the kinetic energy and the potential energy of the EPs. Compared to the change in their kinetic energy, we neglect the contribution from the small change in their potential energy which is proportional to the change in the width of the separatrix (See Appendix C in \cite{Hezaveh2017}). Hence, $E$ can be replaced by $H_{0}$. So we find
\begin{equation}
\dv{E}{t}=\dv{H_0}{t}=\left. \pdv{H_0}{P_1}\right |_{P_2,P_3} G^{\prime}(\Pi)^{-1} \ddot{\alpha},
\label{eq:energyrate}
\end{equation}
where $G=\left. \pdv{H_0}{P_1} \right |_{P_2,P_3}= p\omega_{\tilde{\theta}} + n\omega_{\tilde{\varphi}} =\dot{\alpha}$ previously defined in subsection \ref{subsubsec:Linear} and the last factor on RHS is the rate of chirping of the mode which is the same for all the separatrices corresponding to different $P_2$s in this model. We also have
\begin{equation}
\left. \pdv[2]{H_0}{P_1} \right |_{P_2,P_3} =  \frac{1}{mR_0^2} \left [\frac{p}{q} +n \right ]^2  - \frac{p v_{\parallel} \dv{q}{r} p}{R_0 q^2 \sqrt{2eB_0 P_{\tilde{\theta}}}}.
\label{eq:2}
\end{equation}

The MHD energy of the mode ($E_{\text{wave}}$) is the sum of Eqs. \eqref{eq:1a} and \eqref{eq:1b}, which gives
\begin{eqnarray}
E_{\text{wave}} =& W + K = \frac{1}{2 \mu_{0}}\int \frac{ \dot{\tilde{\bm{A}}}^{2}_{\perp} } {v_{A}^2} + \abs{\nabla \times \tilde{\bm{A}}_{\perp}}^2 \nonumber \\
&- \left (\tilde{\bm{A}}_{\perp} \cdot \nabla \times \tilde{\bm{A}}_{\perp} \right ) \frac{\nabla \times \bm{B} \cdot \bm{B}}{B^2} dV,
\end{eqnarray}
which gives
\begin{equation}
E_{\text{wave}} = 2 \dot{\alpha}^{2} \left [ \bm{\lambda}^{\intercal} \cdot \mathsf{M} \cdot \bm{\lambda} \right ] + 2 \left [ \bm{\lambda}^{\intercal} \cdot \mathsf{N} \cdot \bm{\lambda} \right ]. 
\end{equation}
Substituting the relevant terms into Eq. \eqref{eq:powerbalancecondition1} yields
\begin{eqnarray}
\pdv{\left (\dot{\alpha} - \dot{\alpha}_{t=0} \right )^2}{t} \nonumber \\
=\frac{-8 \gamma_{d} \left [ \dot{\alpha}^{2}  \bm{\lambda}^{\intercal} \cdot \mathsf{M} \cdot \bm{\lambda} + \bm{\lambda}^{\intercal} \cdot \mathsf{N} \cdot \bm{\lambda}     \right ] \left (\dot{\alpha} - \dot{\alpha}_{t=0}\right )   }{ \sum_{P_2} \left [  \iint \delta f dP_{1} d\zeta \frac{\dot{\alpha}}{ \left. \pdv[2]{H_0}{P_1} \right |_{P_2,P_3}}  \Delta P_2\right ]},
\label{eq:sweepingrate}
\end{eqnarray}
where $\dot {\alpha}_{t=0}=\omega_{GAE}$.

\section{Numerical approach}
\label{sec:numerical}

In this section, the numerical approach implemented to solve for the rate of chirping along with the nonlinear mode structure is presented. We have used cubic Hermite elements as the base functions. It is noteworthy that sufficient number of elements should be implemented in order to ensure that the weight/coefficient of each element ($\lambda_l$) varies slowly ($\dv{\ln{\lambda_{l}}}{t} \ll \dot{\alpha}$) during frequency chirping and the radial structure is smooth. The MHD eigenfrequency ($\omega_{\text{GAE}}$) and eigenvector ($\bm{\lambda}_{\text{GAE}}$) of the mode are derived separately by solving the MHD eigenvalue problem by setting the fast particle contribution in Eq. \eqref{eq:generalmodestructure} to zero. The equilibrium profiles used to solve the MHD problem and the resonance condition are given in section \ref{sec:results} (see \fref{fig_equi}).

The general roadmap is as follows

\begin{figure}[!h]
  \centering
\includegraphics[scale=0.5]{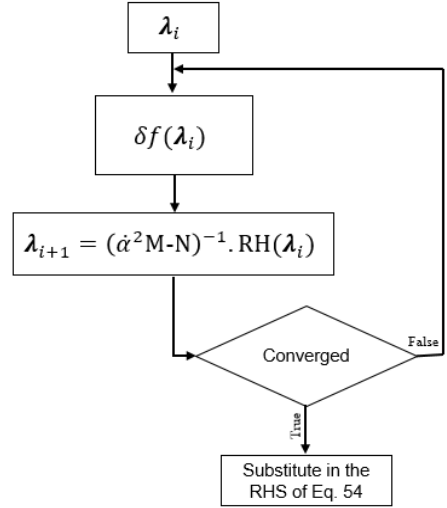}
\caption{The iterative scheme of solving for the mode structure. The index $i$ denotes the number of iteration and RH represents the RHS of Eq. \eqref{eq:generalmodestructure}.}
\label{fig_chart}
\end{figure}

\begin{itemize}
  \item A 5th order Rungge-Kutta method is used to solve the differential equation for the chirping rate
  \item The resonance condition is solved for each new frequency (see section \ref{sec:results})   
  \item At each time step of Rungge-Kutta method, the nonlinear mode structure is calculated iteratively. This stage is visualised in \fref{fig_chart}. 	
\end{itemize}

The explanation of this diagram including the integration over phase-space in RHS of Eq. \eqref{eq:generalmodestructure} or Eq. \eqref{eq:sweepingrate}, how to treat a/an shrinking/expanding separatrix and the special treatment for initial stage is detailed below.

\subsection{Integration over phase-space}

Investigation of the hard nonlinear evolution of the mode structure requires one to consider the contribution of different groups of particles that are simultaneously in resonance with the mode and provide Eq. \eqref{eq:generalmodestructure} with the coressponding pertrubed densities during frequency sweeping. It should be noted that in this model $(P_3=0)$, each $P_2$ corresponds to a slice of resonance line (a specific group of particles in resonance with the mode) associated with a separatrix. For a specific value of $P_2$, there exists a corresponding separatrix in $(P_1,\zeta)$ space whose dynamics affects the mode behavior during chirping. 
\subsubsection{Integration over $P_2$}
The integration over $P_2$ is performed by Trapezoidal rule. As the frequency of the mode begins to deviate from the initial value, there may be some groups of EPs that lose resonance with the mode. On the other hand, there are other groups of EPs whose dynamics satisfy the new resonance condition associated with the updated frequency and will contribute to the interaction. Consequently, after each time step where the frequency is updated, new values of $P_2$ are added to the domain over which the trapezoidal rule is performed.  
\subsubsection{Integration over ($P_1$,$\zeta$)}
Provided that all the separatrices shrink during the evolution, one can integrate over $P_1$ analytically. Nevertheless, it is shown in \cite{Hezaveh2017} that even for a constant trend in frequency sweeping i.e. upward or downward, the value of the adiabatic invariant can have different behaviors depending on the initial equilibrium orbits. Therefore, even for deeply passing energies and for a constant trend in frequency sweeping, the adiabatic invariant corresponding to different groups of resonant particles may exhibit different behaviors in expansion/shrinkage. This needs to be considered in developing a numerical treatment for the evolution of each separatrix \cite{Nyqvist2013,Eremin,wang2012}.
\begin{figure}[t!]
\includegraphics[scale=0.46]{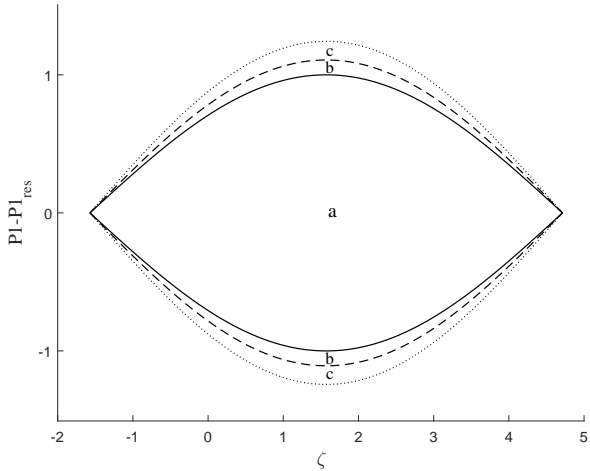}
\caption{Schematic of a separatrix at initial stage (solid line) and expansion of the separatrix after two time steps (dashed and dotted lines). The phase--space area of the initial separatrix is denoted by (a) and the two areas added are specified by (b) and (c).}
\label{separatrix}
\end{figure}
Accordingly, the procedure designed to calculate the integral over each separatrix includes the following main steps:
\begin{itemize}
  \item Calculating the energy related to EPs dynamics on the separatrix for the given $\bm{\lambda}$ and the corresponding value of the adiabatic invariant ($I_{\text{max}}$).
  \item Calculating the ambient phase-space density using the slowing-down distribution given by Eq. \eqref{eq:slowingdown}.

  \item Calculating $\delta$f using Eq. \eqref{eq:slowingdown} and 
	\subitem{- the special treatment presented in \ref{subsec:earlystage} at early stage.}
        \subitem{- the stored phase-space data together with Eq. \eqref{eq:deltaf} during later evolution.}
  \item Storing/updating the phase-space data,
\subitem{-At initial stage: Discretising the separatrix using different adiabatic invariants and assigning an ambient phase-space density value to each region and storing the corresponding data.}
\subitem{-At later evolution: Updating the stored information: Identifying the shrinkage/expansion by comparing each new $I_{\text{max}}$ to the saved data and updating the data accordingly.}
\end{itemize}

The first two steps can be accomplished by using Eq. \eqref{eq:adiabaticinvariant} and the notes thereafter for a constant value of $P_2$. As mentioned above, the numerical scheme should be able to resolve a shrinking as well as an expanding separatrix. In the first instance, this may imply that a fully numerical method should be implemented to perform the integration over phase--space $\left (P_1,\zeta \right )$ since the perturbed phase--space density term can not be taken out of the integral to allow further analytic calculations/simplifications. However, we implement the following justification to further simplify the integral over each region inside the separatrix with a constant value of the distribution function and speedup the calculations: For the case of an expanding separatrix, it is necessary to chirp continuously between the initial and final frequency in order to derive the exact nonlinear structure at a specific frequency after chirping. This means that the corresponding frequency/time step of the numerical approach is chosen to be sufficiently small so that each group of newly trapped particles will carry the value of their distribution function prior to becoming trapped inside the separatrix. Subsequently, sufficiently small time steps result in a sufficiently small phase--space area added around the previous separatrix. Hence, this enables us to consider a flat-top phase-space density over each newly added region/ring around the previous separatrix and simplify the integral for each region with constant density. \Fref{separatrix} shows the phase--space area of an expanding separatrix after two time steps during chirping. Regions b and c represent the small areas added to the initial phase-space area (a) and the value of distribution function is taken to be the same over each region. 

At the initial stage, we discretise the phase-space area surrounded by the separatrix using different values of the adiabatic invariant inside the range $I=[0,I_{\text{max}}]$. This is achieved by substituting different energy values for $K$ in Eq. \eqref{eq:newhamiltonian3}. Each adiabatic invariant is assigned a corresponding value of phase-space density, which represents the value of the distribution function inside the separatrix between two neighbouring discretised adiabatic invariants.  Therefore, we define one adiabatic invariant and one phase-space density vector for each separatrix to track its evolution in the numerical scheme. At each time step, depending on whether the value of the new adiabatic invariant at the separatrix $(I_{\text{max}})$ is greater or smaller than its value at the previous time step, both vectors are being updated.

During the evolution of each phase-space structure, the value of the ambient phase-space density is the same as the equilibrium distribution  function. The difference between this value and the phase-space density inside the separatrix, stored from the previous steps, gives the perturbed distribution function across the separatrix, which can be associated to the height of the coherent structure (hole/clump) in a 3D picture.

\subsection{Early stage of chirping}
\label{subsec:earlystage}

In order to investigate the sweeping rate and the mode structure at early stage of chirping, we rewrite the differential Eq. \eqref{eq:sweepingrate} and Eq. \eqref{eq:generalmodestructure} (with $h=1$) at $t=0$. If the separatrix does not trap new particles (a shrinking separatrix) during chirping, $f_{0}$ remains the same as $F_{0,t=0}$ \cite{Nyqvist2012,Hezaveh2017}. For an expanding separatrix the phase--space density of newly trapped particles should be set to the value of ambient distibution function at the point where the particles are trapped. However, for the very initial stage, one can still set $\delta f=F_0 \left ( P_{1,\text{res},t=0} , P_2 \right )-F_0 \left (P_{1,\text{res}}\left (t \right ), P_2 \right )$. Using the expansion of $F_0 \left (P_{1,\text{res}}\left (t \right ),P_2\right )$ around $t=0$ and $\frac{\Pi-\Pi\left (t=0\right ) }{\dot{\alpha} - \dot{\alpha}_{t=0}}\approx\left. \left (\pdv[2]{H_0}{P_1} \right )^{-1} \right |_{P_1=\Pi\left (t=0\right )}$, we find
\begin{eqnarray}
\pdv{\left (\dot{\alpha} - \dot{\alpha}_{t=0} \right )^2}{t} = \nonumber \\
\frac{8 \gamma_{d} \left [ \dot{\alpha}^{2} \bm{\lambda}^{\intercal} \cdot \mathsf{M} \cdot \bm{\lambda} + \bm{\lambda}^{\intercal} \cdot \mathsf{N} \cdot \bm{\lambda}\right ]}{ \sum_{P_2}  \left [\iint  dP_{1} d\zeta \pdv{F_0}{P_1} \left.  \right |_{P_1=\Pi\left (t=0\right )} \frac{\dot{\alpha}}{  \left (\pdv[2]{H_0}{P_1} \right )^2 \left. \right |_{P_1=\Pi\left (t=0\right )}}  \Delta P_2 \right ]}
\label{eq:initialsweepingrate}
\end{eqnarray}
at initial stage of frequency chirping. For analysing the saturated mode structure at early stage of chirping, we write $\dot{\alpha}= \dot{\alpha}_{t=0} + \Delta\dot{\alpha}$ and substitute in Eq. \eqref{eq:generalmodestructure} (with $h=1$) for $\dot{\alpha}$, to have
\begin{eqnarray}
a = \frac{\bm{\lambda}_{\text{GAE}}^{\intercal}}{8\dot{\alpha}\bm{\lambda}_{\text{GAE}}^{\intercal} \cdot \mathsf{M} \cdot \bm{\lambda}_{\text{GAE}}} \int d^3p d^3q \pdv{F_0}{P_1}   \nonumber \\
\times \left (\pdv[2]{H_0}{P_1} \right )^{-1} \left. \right |_{P_1=\Pi\left (t=0\right )} \sum_p \begin{bmatrix} V_{p,n,l=1} \\ \vdots \\  V_{p,n,l=\textbf{s}} \end{bmatrix} \e^{i\zeta_{p}  } + c.c.
\label{}
\end{eqnarray}
where we have considered the saturated mode struture to be a linear factor of the MHD eigenvector ($\bm{\lambda}_{\text{GAE}}$), $\bm{\lambda}=a\bm{\lambda}_{\text{GAE}}$. For an eigenmode growing outside the shear \Alfven continuum, the structure of the radial profile remains almost the same as the initial eigenvector\cite{Berk1995}.  
 
\section{Results}
\label{sec:results}

We set the values of physical parameters as follows: the axial magnetic field at center $B_{\varphi}(r=0)=\SI{2}{\tesla}$, $R_0=3.5\si{\meter}$, the minor radius $r_m=1\si{\meter}$, the ion mass $m_i\approx  3.3\times 10^{-27}\si{\kilogram}$, the number density of bulk plasma ions $n_{\text{Bulk}}=5\times 10^{20}\si[inter-unit-product =$\cdot$]{\per\cubic\meter}$. The fast particles density $n_f$ is taken to be $ (1-10)\% $ of $n_{\text{Bulk}}$.

\subsection{Equilibrium profiles and resonance condition}
\label{subsec:}
\begin{figure}[!t]
  \centering
\includegraphics[scale=0.5]{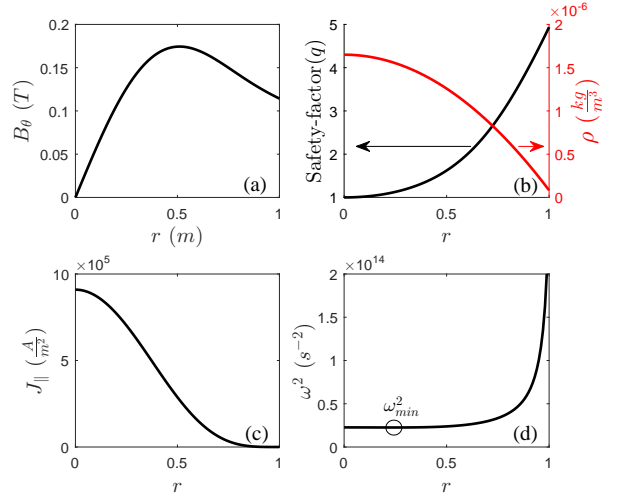}
\caption{MHD equilibrium profiles: (a) the poloidal component of the magnetic field (b) the black and red lines represent the q and the density profiles, respectively (c) the parallel current density and (d) \Alfven continuum for $m=3$ and $n=9$.}
\label{fig_equi}
\end{figure}

\begin{figure}[!t]
  \centering
\includegraphics[scale=0.5]{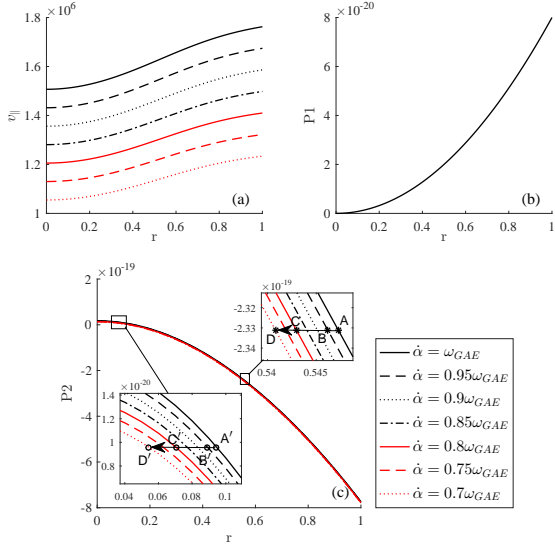}
\caption{Resonance line of $v_{\parallel}$ (a), $P_1$ (b) and $P_2$ (c) for different wave frequencies. Points $A$ and $A^{\prime}$ on (c) represent the initial position of a separatrix that moves to points $\{B,C,D\}$ and $\{B^{\prime},C^{\prime},D^{\prime}\}$ , respectively, during frequency chirping while the value of $P_2$ is preserved.}
\label{fig_fastpara} 
\end{figure}

For the purpose of this work, we consider the density and current profiles mentioned in \cite{Appert} to solve the equilibrium condition and the MHD eigenmode problem (Eq. \eqref{eq:generalmodestructure} without the contribution of fast particles) in the low-$\beta$ limit. \Fref{fig_equi} shows the equilibrium parameters as a function of radial position. It should be mentioned that we have benchmarked the eigenvectors of the MHD eigenvalue problem code with the radial component of displacement vectors reported in \cite{Appert}. In this model, the GAE exists just below the shear \Alfven continuum since the singularity in the eigenfunction no longer occurs at $\left. r \approx r \right |_{\omega=\omega_{\text{min}}} $ due to the inclusion of the current dependent terms. The choice of the mode numbers is based on two factors: An MHD eigenmode should exist for the corresponding mode numbers and also there should be sufficient drive for the corresponding mode with respect to a realistic description of EPs distribution. Consequently, we have considered the mode numbers $m=3$, $n=9$ for the case of a slowing down distribution of energetic ions presented by
\begin{equation}
F_0 = \frac{n_0A}{v_\parallel^3 + v_c^3} \e^{\frac{P_{\varphi}}{\Delta P_{\varphi}}}\delta \left ( P_3 - 0^{+} \right )
\label{eq:slowingdown}
\end{equation}
where $n_0$ is the density of the fast particles at the center, $A=\frac{3\sqrt{3} v_c^2}{4\pi^2 e B m_i}$ is the normalization constant with $v_c$ the critical velocity and $\Delta P_{\varphi}$ the width of $F_0$ on $P_{\varphi}$. The aforementioned mode numbers correspond to an eigenfrequency of $\omega_{\text{GAE}}=4.73\times10^{6}\si{\radian \per \second}$  where radial wavenumber is 1. The fast ions parameters are chosen to satisfy $\gamma_l \ll \omega_{\text{GAE}}$. By setting $v_c=2.6\times10^{6}$, $\Delta P_{\varphi}=0.47 \times 10^{-20}$ and $n_f=10\%n_{\text{Bulk}}$, we find the linear growth rate $ \gamma_l= 1.13\times10^{4}\si{\per\second}$ for $p=2$. It is noteworthy that for highly co-passing energetic ions studied in this model, the coupling strength is nonzero for $p=m \pm 1$ (see \ref{App1}). Prior to investigating the evolution of the nonlinear structure, the resonance condition Eq. \eqref{eq:resonance} should be solved to find the values of action variables, Eq. \eqref{eq:CT1}, at the resonance and also to track the dynamics of the resonant particles. For a downward trend in frequency chirping, \fref{fig_fastpara}a, b and c illustrate the resonance line for $v_{\parallel}$, $P_1$ and $P_2$ respectively, versus radial position for different frequencies. The conservation of conjugate momenta $P_2$ and $P_3$, with the latter being $\approx 0$ for deeply passing EPs, allows us to track the dynamics of EPs and identify whether the motion of nonlinear structures (holes/clumps) in phase--space results in an inward or outward flux of the fast ions in resonance with the chirping GAE. 
Since $P_2$ corresponds to an ignorable coordinate, its value must be preserved during the motion of holes/clumps. Therefore, the motion of the corresponding separatrix in the radial direction occurs in a way that the value of $P_2$ remains the same during chirping. As an example, the group of particles that satisfy the resonance condition at point A on \fref{fig_fastpara}c should move to point D in order to conserve the value of $P_2$ while satisfying the resonance condition. Therefore, this results in an inward flux of fast ions towards the plasma core during frequency sweeping of the eigenmode.

\begin{figure}[!t]
  \centering
   \includegraphics[scale=0.58]{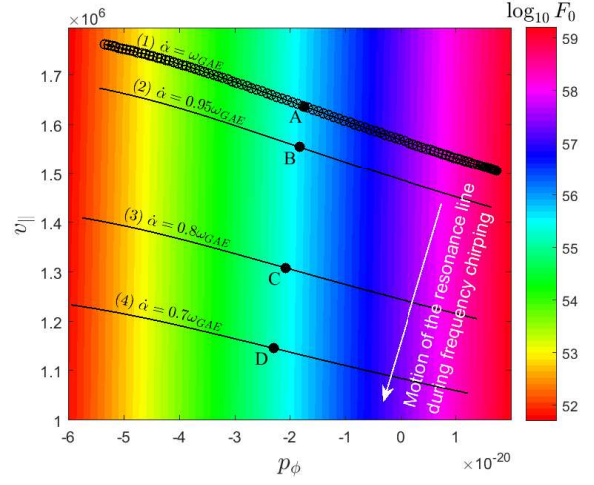}
  \caption{The equilibrium phase-space density and the resonance line at different frequencies. The initial resonance line is denoted by unfilled circles implying that the phase--space density of each separatrix is the same as the equilibrium distribution function shown at the background. Points A,B,C and D correspond to the position of the same separatrix on both \fref{fig_fastpara}c and \fref{fig_phase_space} at each corresponding frequency, e.g. Point B denotes the radial and the phase-space position of the same separatrix on \fref{fig_fastpara}c and \fref{fig_phase_space}, respectively at $\dot{\alpha}=0.95\omega_{\text{GAE}}$.}
\label{fig_phase_space}
\end{figure}

\begin{figure}[!t]
  \centering
   \includegraphics[height=6.5cm, width=8.5cm]{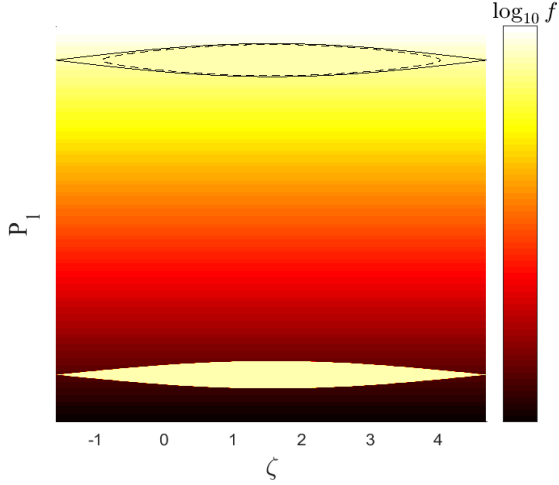}
  \caption{Evolution of the phase--space clump. The dashed line represents the adiabatic invariant corresponding to the separatrix at the bottom.}
  \label{fig_clump}
\end{figure}

\Fref{fig_phase_space} shows the equilibrium phase--space density of EPs as well as the resonance line for four different frequencies. Initially, the value of the total phase-space density inside each separatrix is the same as the equilibrium distribution function. This is illustrated by using circles on the initial resonance line (1) in \fref{fig_phase_space}. As the frequency chirps, the separatrices (phase--space structures) preserve the initial value of the phase-space density during their motion and carry the initially-in-resonance EPs with the mode to new regions in phase-space. Therefore, depending on whether the value of $F_0$ at these new regions are lower or higher than the value of $F_0$ at initial resonance i.e. $\delta f>0$ or $\delta f<0$ (see Eq. \eqref{eq:perturbedpart}), a clump or hole will be developed inside the separatrix. 
It should be mentioned that if the separatrices trap new EPs on their way due to expansion (see \fref{separatrix}) and carry them, the preservation of the phase--space density of the newly trapped EPs should also be taken into account. The separatrix, which is initially located at point A on \fref{fig_phase_space}, will move to points B, C and D at each corresponding frequency. It can be observed that the separatrices move to regions where the value of the ambient equilibrium phase-space density is lower. Therefore, a clump will be developed inside the phase-space structure to preserve the distribution function value inside the separatrix. \Fref{fig_clump} illustrates the total distribution function inside the separatrix at point A which moves down in $P_1$ to point B as a result of frequency chirping in this model.        

Further explanation can be given to identify the phase-space structures as clumps: In this case, the value of $\pdv{F_0}{P_1}|_{P_1=P_{1,\text{res}}}$, namely the drive, remains positive for all the separatrices during frequency chirping of the mode. For deeply passing particles, we have
\begin{equation}
X_r=r_0+\Delta r\cos\theta,
\label{eq:Xr}
\end{equation}         
and using Eqs. \eqref{eq:ptheta2} and \eqref{eq:CT1}, we find
\begin{equation}
P_{\tilde{\theta}}=\frac{1}{2}eB_0 r_0^2.
\label{eq:pthetatilda}
\end{equation}
In this case, the separatrices move inward in the radial direction, hence the value of $P_{1,res}$ for each separatrix decreases as the mode chirps down as does the value of the ambient distribution function. This means that in this case, the phase-space structures are clumps.

An upward trend of frequency sweeping in this case will result in an outward flux of the fast ions towards the first wall and the phase--space coherent structures will be holes. However, it should be taken into consideration that the model remains valid only for eigenmodes subject to weak continuum damping where the linear structure is not mainly identified by the fast particles as opposed to energetic particle modes (EPMs). Therefore, crossing the continuum edge should be avoided during chirping.

\subsection{chirping rate, structure evolution and adiabaticity validation}
\label{subsec:}

For the case of a near threshold instability $\abs {\gamma_l-\gamma_{d}} \ll \gamma_{d} \le \gamma_l$, we choose $\gamma_{d}=1.1\times10^{4}\si{\per\second}$ and solve differential Eq. \eqref{eq:sweepingrate} coupled with the integral Eq. \eqref{eq:generalmodestructure} with the approach mentioned in section \ref{sec:numerical} to determine the nonlinear behavior during long range frequency chirping. 

\begin{figure}[!t]
  \centering
   \includegraphics[scale=0.7]{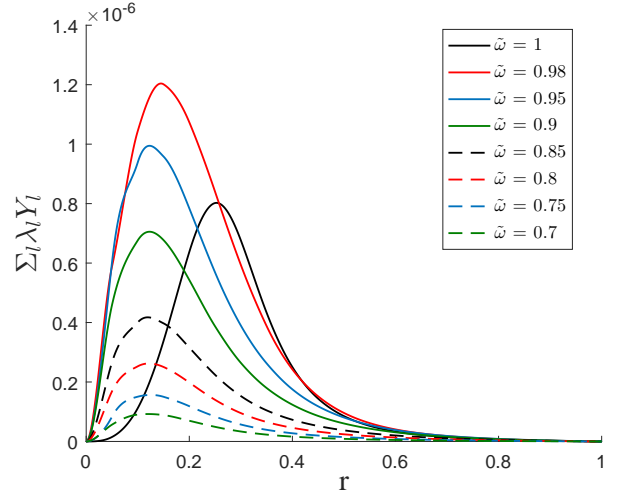}
  \caption{Evolution of the radial profile of the GAE during frequency chirping. $\tilde{\omega}=1$ represents no change in the mode frequency}
\label{fig_structure}
\end{figure}

The radial current created by the population of the energetic ions modifies the structure of the MHD eigenmode in the hard nonlinear regime. \Fref{fig_structure} demonstrates the evolution of the radial profile of $\phi\left (\bm{r},t\right )$ while the frequency of the mode deviates from the initial eigenfrequency. The peak of the initial eigenmode structure, located at the point where the extremum of the \Alfven continuum occurs, will be shifted inward towards the center of the plasma and the mode becomes more localized close to the plasma center. This inward displacement is in compliance with the inward drift of EPs explained above.
In addition, it can be observed that the radial profile is broadened as the frequency moves away from the shear \Alfven continuum. As the frequency decreases, the amplitude of the radial profile initially grows and then starts to decrease. It is noteworthy that the amplitude value at $\tilde{\omega}=1$ represents the saturated amplitude corresponding to the aforementioned linear growth rate. In this model, the axial current resolves a pole in the MHD equations and allows weakly damped GAEs with smooth radial profiles as opposed to highly damped continuum modes with spiky radial structures. However, the eigenfrequency of this GAE lies just below the shear \Alfven continuum and the initial frequency is very close to the value corresponding to the pole in the MHD equations. Therefore, for a fixed frequency change, the mode structure changes more when the frequency change occurs closer to the initial eigenfrequency. This can be investigated using \fref{fig_structure}. It has been shown that the radial profile changes more when the frequency changes from $\tilde{\omega}=1$ to $\tilde{\omega}=0.98$ as opposed to the case where it changes from $\tilde{\omega}=0.98$ to $\tilde{\omega}=0.95$.  

The change in the radial component of various plasma quantities during chirping can also be analyzed using $\Phi \left (\bm{r},t\right )$. The displacement vector reads,
\begin{equation}
\bm{\xi} = - \frac{1}{B_0^2} \left (\tilde{\bm{A}}_{\perp} \times \bm{B}_0 \right ). 
\label{eq:displacement1}
\end{equation}
Using Eq. \eqref{eq:A1}, we have
\begin{eqnarray} 
\bm{\xi}B_0 =& \pdv{\Phi \left (\bm{r},t\right )}{r} \hat{\bm{e}}_{\perp} - \nonumber \\
&i\frac{m}{r}\left (\Phi \left (r,t\right ) \e^{i\left (m\theta + n\varphi - \alpha \left (t \right )\right)} - c.c \right ) \hat{\bm{e}}_r,
\label{eq:displacement2}
\end{eqnarray}
where the poloidal component of the equilibrium magnetic field has been neglected compared to the toroidal component. 
\begin{figure}[!t]
  \centering
   \includegraphics[scale=0.6]{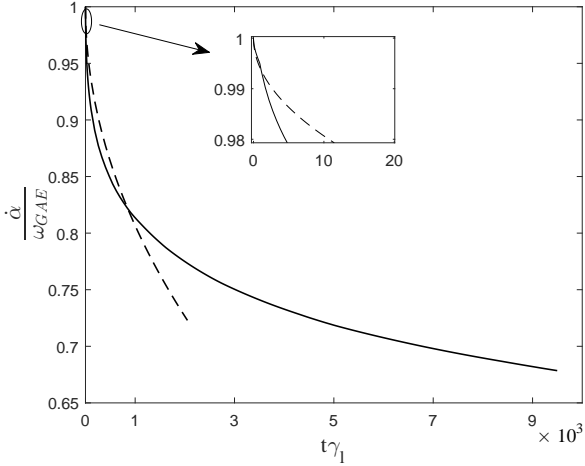}
  \caption{The solid line represents the time evolution of the mode frequency in the hard nonlinear regime and the dashed line, plotted for comparison, corresponds to the square root dependency.}
  \label{fig_chirping_rate}
\end{figure}
Eq. \eqref{eq:displacement2} clearly demonstrates the relation between radial component of the displacement vector and the radial mode structure plotted in \fref{fig_structure}a. \Fref{fig_chirping_rate} illustrates the rate at which the frequency chirps. It is shown that the square root dependency holds for the very early stages of frequency chirping. 

The adiabatic condition represented in subsection \ref{subsubsec:Chirping}, which is implemented for the formalism, needs to be validated if it remains satisfied \cite{Berk1999,Eremin,wang2012NF}. Eq. \eqref{eq:newhamiltonian3} can be written as
\begin{eqnarray}
K=\frac{1}{2}\pdv[2]{H_0}{P_1} & \left (\Pi, P_2, P_3 \right )     \left [ P_1 - \Pi  \right ]^2 \nonumber \\
&-\sum_{l} 2\abs{\lambda_l V_{p, n, l}} \cos \left (\zeta + \sigma \right ),
\label{eq:simpHamiltonian_adb}
\end{eqnarray}
where $\sigma=\tan^{-1}\frac{\Im(\lambda_l V_{p, n, l})}{\Re(\lambda_l V_{p, n, l})}$ and for the case of EPs with highly passing orbit types, we have $\sigma=\pm \frac{\pi}{2}$. It is worth noting that in this case we have $\Re(\lambda_l V_{p, n, l})=0$. Using canonical equations of motion, one finds
\begin{align}
&\dot{P_1}  =- \sum_l 2\abs{\lambda_l V_{p, n, l}} \cos \left (\zeta + \sigma \right ),	\refstepcounter{equation} \label{eq:canonicaleq1} \subeqn \\
&\dot{\zeta} = \pdv[2]{H_0}{P_1} \left [ P_1 - \Pi  \right ]^2.  \label{eq:canonicaleq2}  \subeqn
\end{align}
The motion of deeply trapped EPs inside the separatrix satisfies the pendulum equation
\begin{eqnarray}
\dv[2]{}{t}\left (\zeta+\sigma \right)=&-\pdv[2]{H_0}{P_1} \sum_l 2\abs{\lambda_l V_{p, n, l}} \nonumber \\
&\times \sin \left (\zeta+\sigma\right ),
\label{eq:}
\end{eqnarray}
\begin{figure}[!t]
  \centering
   \includegraphics[scale=0.65]{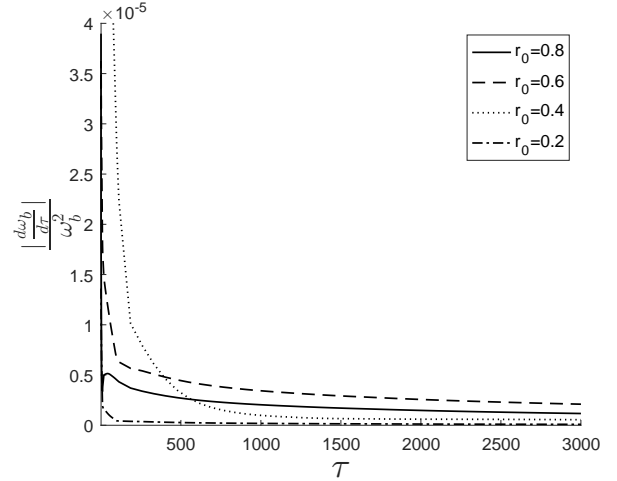}
  \caption{The evolution of time rate of change in the bounce frequency divided by squared value of the bounce frequency for different initial radial positions ($r_0$) of the separatrices i.e. different values of $P_{1,res}$ before chirping.}
\label{adiabatic}
\end{figure}
where we have used $\sin\left(\zeta+\sigma\right) \approx \left (\zeta+\sigma\right)$ at the center of the separatrix, the so-called O-point. As shown in \fref{separatrix}, we have $\sigma=-\frac{\pi}{2}$ for the results reported in this paper and the O-point is located at $\zeta=\frac{\pi}{2}$. Subsequently, the bounce frequency of deeply trapped EPs inside the separatrix is
\begin{equation}
\omega_b = \sqrt{2\pdv[2]{H_0}{P_1} \sum_l \abs{\lambda_l V_{p, n, l}}}.
\label{eq:bouce}
\end{equation}
The RHS of adiabatic condition $1 \gg \frac{\abs{\dv{\omega_b}{t}}}{\omega_b^2}$ is plotted in \fref{adiabatic} for separatrices with different initial radial positions. Consistent with the previousely reported results \cite{wang2012NF,Hezaveh2017}, we also observe that the adiabatic condition is never formally satisfied at the very early stage of chirping. However, it is shown that once the adiabatic condition is satisfied, it remains valid for later evolution of the mode. In addition, it was shown in \cite{Hezaveh2017} that the assumption of $\gamma_l \ll \omega_{\text{GAE}}$ implies that the period during which the adiabatic limit is not satisfied is very short.        

\section{Concluding remarks}
\label{sec:conclusion}

A theoretical description has been developed to study the hard nonlinear evolution of a global \Alfven eigenmode (GAE) in resonance with co-passing energetic particles (EPs) in an NBI scenario during the adiabatic frequency chirping behavior of the mode. Constructing appropriate constants of motion allows us to track the dynamics of EPs as the frequency of the mode changes. In addition, a finite element method using cubic Hermite base functions, has been implemented to represent the radial profile of the GAE. This enables the derivation of an analytic expression for the nonlinear radial structure of the mode by varying the total Lagrangian of the system with respect to the weight of the finite elements. Hence, the radial structure can be updated as the frequency deviates from the initial MHD eigenfrequency. During chirping, the possibility of both the shrinkage and the expansion of a separatrix has been taken into account since different separatrices in phase-space may exhibit different behaviors. Phase-space structures, identified to be clumps, move in order to extract energy from the EPs distribution function and deposit it into the bulk plasma. This energy balance is used to derive an expression for the time rate of change in the frequency. The adiabatic condition is also evaluated which remains valid once it is satisfied.     

Energetic-ion parameters, such as orbit width or pressure, can cause a shift and a broadening in the radial profile of the mode \cite{Todo}. In addition, for the case of a near threshold instability, we have shown how the deviation of the frequency from the initial eigenfrequency can also result in shifting the peak location of the radial profile and also radial broadening during the hard nonlinear stage. For the case presented in this manuscript, the slowing down EPs move radially inward as clumps when the frequency chirps downward. The orbit width of the EPs follows
\begin{equation}
\Delta r=\frac{q m_i v_\parallel}{eB},
\end{equation}
where $q$ is the safety factor, $m_i$ is the ion mass, $v_{\parallel}$ is the velocity of EPs parallel to the equilibrium magnetic field, $e$ is the electron charge and $B$ is the magnetic field. The range of orbit widths from plasma center to the boundary is $0.01$-$0.08\si{\meter}$. This corresponds to an energy range of 23-32 kev for EPs initially in resonance with the mode.   
            
A comprehensive description of the problem is aimed in our research plan. This includes 
\begin{itemize}
\item adding the toroidal effects to the bulk description,
\item allowing the EPs nonlinearity to simultaneously update all the components of the mode structure, namely poloidal, toroidal and radial and
\item allowing the frequency to cross the continuum edge and behave as realistically as possible inside the continuum.
\end{itemize}
In this manuscript, we have taken a short-cut, associated with some assumptions, to the above roadmap to build the presented model along the way and set a feeling of the full problem which requires more severe effort but feasible. In addition, the calculation of the EPs dynamics is done for deeply passing particles where the coupling strength is nonzero for only two values of $m$, i.e. $m\pm 1$. A further study is to investigate the interaction when EPs are magnetically trapped and follow the so-called banana orbits.

\section*{Acknowledgments}
This work was funded by the Australian Research Council through Grant No. DP140100790. The first author is very grateful to the Australian Nuclear Science and Technology Organisation (ANSTO) and the Festival de Th\'eorie for travel supports to present this work.   

\section*{References}

\bibliography{references}
\bibliographystyle{iopart-num}

\clearpage

\appendix

\section{Calculation of the coupling strength}
\label{App1}

Using Eq. \eqref{eq:phidot} and the resonance condition ($ p \dot{\tilde{\theta}} = \dot{\alpha} \left ( t \right ) + n \dot{\tilde{\varphi}} $), Eq. \eqref{eq:V_p} can be written as
\begin{eqnarray}
V_{p,n,l} = & \frac{1}{2 \pi} \int_{0}^{2 \pi} ie \left [ p \dot{\tilde{\theta}} - \frac{m \dot{\varphi}}{q \left ( X_r \right )} +  n \dot{\tilde{\theta}} \pdv{\Delta \varphi}{\tilde{\theta} }  \right ] \nonumber \\
& \times Y_l \left ( X_r \right )  \e^{im \theta - in \Delta \varphi - ip \tilde{\theta}} d \tilde{\theta}.
\label{eq:V_pp}
\end{eqnarray}
The integral over the first term of the integrant can be performed using integration by parts. Therefore, 
\begin{eqnarray}
V_{p,n,l} =& \frac{1}{2 \pi} \int_{0}^{2 \pi} e \left [  \dot{\tilde{\theta}} \pdv{X_r}{\tilde{\theta}} \dv{Y_l \left ( X_r \right )}{r} + im \left (   \dot{\theta} - \frac{\dot{\varphi}}{q \left (X_r \right)}  \right ) \right. \nonumber \\ 
& \left. \times  Y_l \left ( X_r \right ) \right ] \e^{im \theta - in \Delta \varphi - ip \tilde{\theta}} d \tilde {\theta}.
\end{eqnarray}
Eqs. \eqref{eq:ptheta2} and \eqref{eq:pthetadot} can be used to find
\begin{equation}
\dot{X}_r = \dot{\tilde{\theta}} \pdv{X_r}{\tilde{\theta}} =  - \left [ \frac{m_i v_{\parallel}^2 }{R} +\frac{\mu B_0}{R_0} \right ] \sin \theta \frac{1}{e B_0}.
\end{equation}
Simple implementation of Eqs. \eqref{eq:phidot} and \eqref{eq:thetadot} gives
\begin{equation}
\dot{\theta} - \frac{\dot{\varphi}}{q \left(X_r \right)} = - \left [ \frac{m_i v_{\parallel}^2 }{R} +\frac{\mu B_0}{R_0} \right ] \cos \theta \frac{1}{e B_0 r_0}.
\end{equation}
Under the small orbit width assumption, $V_{p,n,l}$ reads
\begin{eqnarray}
V_{p,n,l} =& - \frac{1}{2 \pi} \int_{0}^{2 \pi} \left [ \frac{m_i v_{\parallel}^2 }{R B_0} +\frac{\mu}{R_0} \right ] \left [ \dv{Y_l \left ( r_0 \right ) }{r}  \sin \theta \right. \nonumber \\
& \left. + \frac{i m Y_l \left ( r_0 \right ) }{r_0} \cos \theta \right ] \e^{im \theta - in \Delta \varphi -ip \tilde{\theta}} d \tilde{\theta}.
\label{eq:finalv}
\end{eqnarray}
For deeply passing EPs inside the equilibrium field, one can neglect the infinitesimal perpendicular velocity of the particles to the magnetic field and set $\mu \approx 0$. In this limit, $v_{\parallel}$ becomes a constant of motion and we can set $\theta \approx \tilde{\theta}$ and $\Delta \varphi=cte$. Using Euler's formula and the orthogonality of trigonometric functions, one finds
\begin{equation}
V_{p,n,l}=\frac{-i m_i v_\parallel ^2}{2 B_0 R_0} \left [\pm Y_l^{\prime}  + \frac{m}{r_0} Y_l \right ], p=\left ( m \mp 1 \right )
\end{equation}
It should be noted that we have set $\Delta \varphi=0$. Non-zero values of $\Delta \varphi$ results in a shift of the separatrix in phase--space compared to the existing model.

\end{document}